# A phenomenological dislocation mobility law for bcc metals

Giacomo Po[a], Yinan Cui[a], David Rivera[a], David Cereceda[b,c], Tom D. Swinburne[d], Jaime Marian[c], and Nasr Ghoniem[a,c]

[a]*Mechanical and Aerospace Engineering Department, University of California, Los Angeles, CA 90095, USA*
[b]*Hopkins Extreme Materials Institute, Johns Hopkins University, Baltimore, MD 21218, USA*
[c]*Materials Science and Engineering Department, University of California, Los Angeles, CA 90095, USA*
[d]*CCFE, Culham Science Centre, Abingdon, Oxon, OX14 3DB, UK*

**Abstract**

Dislocation motion in body centered cubic (bcc) metals displays a number of specific features that result in a strong temperature dependence of the flow stress, and in shear deformation asymmetries relative to the loading direction as well as crystal orientation. Here we develop a generalized dislocation mobility law in bcc metals, and demonstrate its use in discrete Dislocation Dynamics (DD) simulations of plastic flow in tungsten (W) micro pillars. We present the theoretical background for dislocation mobility as a motivating basis for the developed law. Analytical theory, molecular dynamics (MD) simulations, and experimental data are used to construct a general phenomenological description. The usefulness of the mobility law is demonstrated through its application to modeling the plastic deformation of W micro pillars. The model is consistent with experimental observations of temperature and orientation dependence of the flow stress and the corresponding dislocation microstructure.

*Keywords:* dislocation mobility, bcc metals, non-Schmid effects,, tungsten

## 1. Introduction

Dislocation mobility is a fundamental property of crystals that determines many characteristics of their plastic deformation. In computational multiscale methods of plastic deformation (e.g. discrete or continuum DD, crystal plasticity, continuum plasticity), dislocation mobility is modeled by a *phenomenological* law, which prescribes the dependence of the steady-state velocity of a dislocation on local parameters, such as line character, slip system, stress, temperature, and composition.

In fcc crystals, the task of formulating mobility laws is simplified by several features of close-packed {111}⟨110⟩ slip. First, the lattice resistance to dislocation motion is typically negligible compared to the applied force, ultimately as a consequence of the planar character of the disassociated dislocation core and the reduced magnitude of the partial Burgers vector [1, 2]. Second, the planar core structure results in dislocation glide being controlled by the resolved shear stress only. Thus, dislocation motion obeys closely Schmid's postulation [3]. Third, in lieu of Neumann's principle [4], the mirror symmetry of the fcc lattice with respect to {110} planes (i.e. planes normal to the slip direction) guarantees that the flow stress is independent of slip plane orientation and sense of shear (fig. 1). As a consequence, dislocation glide can be considered an "atomically smooth" process, and simple mobility laws can be obtained from the condition that the mechanical force balances the dissipative friction force. In particular, for small dislocation velocity compared to the shear wave speed, the interaction of phonons with dislocations provides the main mechanism of energy dissipation, resulting in a drag force that depends linearly on the dislocation speed. The type of mobility law that emerges from these considerations typically involve a drag coefficient, which may depend on temperature and dislocation character (screw vs edge).

It bears emphasis that fcc-type mobility laws automatically satisfy Schmid's law, the empirical yield criterion formulated in the pre-dislocation era by Schmid [3, 5] for close-packed metals subject to uniaxial loads. The law contains two distinct statements [6]. First, that plastic deformation initiates on the (low-index) crystallographic plane with the highest Schmid factor, when the stress resolved on that plane in the slip direction reaches a critical value. Second, that the critical value is a *constant* material parameter, known as the *critical resolved shear stress* ($\tau_{\text{crss}}$). In particular, the $\tau_{\text{crss}}$ does not depend on orientation of the slip plane, sense of the load, and local stress state.

Owing to the pioneering work of Taylor [7, 8], it was soon recognized that bcc metals do not strictly obey Schmid's law. Early studies [9, 10] revealed remarkable features common to several bcc transition metals, such as the *twinning/anti-twinning* (TAT) and the *tension/compression* (TC) asymmetries of the yield and flow stresses, macroscopically non-crystallographic slip, and a strong temperature and strain rate dependance of the flow stress.

Hirsch [11] first identified the origin of these features in the non-planar character of the core of screw dislocations, whose stress-free configuration spreads equally on equivalent low-index planes of the ⟨111⟩ zone. Several elastic, atomistic, and *ab-initio* studies followed this rationalization and shed light on the structure of screw dislocation cores in bcc metals [see 6, 12, 13, 14, for a review]. Early atomistic simulations predicted an extended and degenerate (or polarized) core structure for group VB metals, while a compact and non-degenerate core was found for group VIB metals [4]. More recent electronic structure calculations, however, support the evidence that pure transition metals posses a compact and non-degenerate core structure [15]. The two most prominent aspects that stem from this particular core structure are [16, 17]:

- *The temperature-dependence of the flow stress.* Because motion requires a modification of the non-planar core configuration, screw dislocations experience a high lattice resistance. As a consequence, below a certain temperature-dependent critical stress, glide of screw dislocations proceeds by way of thermally-assisted nucleation and subsequent lateral migration of kinks-pairs [18]. The rate of these two processes ultimately controls the velocity of the screw dislocation as a whole. Because kink nucleation is a thermally-activated process, the flow stress rapidly increases with decreasing temperature. Moreover, the ability of screw dislocations to proceed by kinks on different planes of the $\langle 111 \rangle$ zone may manifest in macroscopically non-planar slip, with surface slip traces showing a *pencil glide* pattern when observed along the slip direction [7, 12].

- *The emergence of non-Schmid effects.* There are deviations from both statements contained in Schmid law, which in general change the condition for the onset of slip into a function of stress components other than the resolved shear stress. Originally undistinguished, two different phenomena are now recognized [4, 19, 16], namely the TAT asymmetry, and the TC asymmetry of the flow stress. The TAT asymmetry is an intrinsic property of the bcc lattice related to its lack of mirror symmetry with respect to planes orthogonal to the dominant $\langle 111 \rangle$ slip direction (fig. 1). On the other hand, the TC asymmetry is due to components of the local stress other than the glide component, which influence the structure of the screw core and thus the critical conditions for dislocation motion.

Several phenomenological mobility laws have been proposed to take into account these important features, and have been used in DD simulations. Several authors [20, 21, 22] have implemented a temperature-dependent screw dislocation mobility by accounting for the thermally-activated kink-pair formation process. In similar settings, Chaussidon et al. [23] have considered a cross-slip probability influenced by the TAT orientation of the slip plane. Wang et al. [24] have proposed a mobility law with different values of the Peierls stress for [111]{112} twinning and anti-twinning dislocations. Srivastava et al. [25] account for both TAT and TC asymmetries by supplying the kink-pair activation enthalpy as a function of orientation as a lookup table compiled from atomistic calculations.

The objective of the present work is to develop an analytical mobility law for dislocation motion so as to reproduce observations and phenomena associated with plastic flow in bcc metals. The aim is to establish a procedure for accurate, yet computationally convenient description of dislocation velocity, where the influence of the applied stress state, temperature, and possibly alloying elements can be incorporated. The intended utilization of the law is in large-scale DD simulations, and in dislocation-based crystal plasticity models. To achieve our goal, we will use fundamental theory, together with experimental measurements and atomistic computer simulations (Molecular Statics, Molecular Dynamics, and First Principles) to establish a *phenomenological* law for dislocation mobility. The purpose of the theory, presented in section 2, is to guide the development of an analytical form, inspire the phenomenological description of dislocation motion, and to furnish an understanding of its physical origins. As we shall see later, the proposed law, developed in section 3 has fitting parameters obtained from detailed computer simulations, and calibrated with experimental data. We describe a *process* by which mobility parameters are established, and demonstrate the applicability of the developed mobility law in DD simulations. Since the main goal here is to allow efficient DD simulations of bcc metals, we present an application of the developed law to the study of temperature effects on the dislocation microstructure and flow stress in W micro-pillars in section 4. Finally, conclusions of the work and future directions are given in section 5.

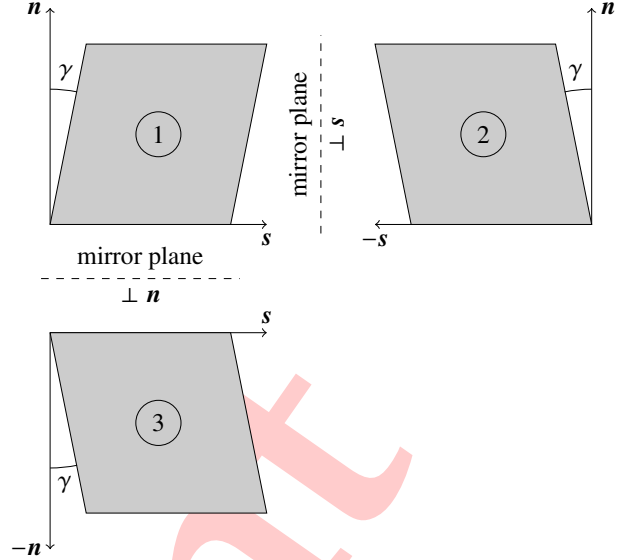

Figure 1: Sketch of Neumann's principle in relation to the symmetry of the flow stress [26]. Crystal (1) is plastically deformed in shear with distortion $\beta_1 = \gamma s \otimes n$. Crystal (2) is deformed with opposite distortion $\beta_2 = \gamma(-s) \otimes n$, by shearing it along the opposite slip direction. If the two crystals lattices are mirror-symmetric with respect to planes perpendicular to $s$, then the two deformed crystals are indistinguishable. This symmetry guarantees the independence of the flow stress on the sense of shear, as in the case of $\langle 110 \rangle$ slip in fcc crystals, for any slip plane of the $\langle 110 \rangle$ zone. It does *not* apply to $\langle 111 \rangle$ slip in bcc metals. Crystal (3) is also subject to the opposite distortion $\beta_3 = \gamma s \otimes (-n)$. If the crystals lattices (1) and (3) are mirror-symmetric with respect to planes perpendicular to $n$, then the two deformed crystals are indistinguishable from each other. This symmetry applies to {110} planes in bcc metals. Therefore $\langle 111 \rangle${110} slip should be unaffected by the TAT asymmetry. However, recent studies [27, 28] have shown that the actual trajectory of $\langle 111 \rangle${110} screw dislocations between consecutive "easy-core" configurations bows out of {110} planes.

## 2. Background theory

The Peierls stress $\tau_p$ is defined as the minimum applied shear stress needed at 0 K to move an infinitely long dislocation over the periodic misfit energy barrier of the glide plane at 0 K [29, 30]. $\tau_p$ is strongly dependent on the crystal lattice structure, basis atoms, and slip system. The magnitude of $\tau_p$ spans the broad range $10^{-5} - 10^{-1}\mu$ for most crystalline materials, where $\mu$ is the shear modulus [31]. Values of $\tau_p$ in fcc metals for both edge and screw $\langle 110\rangle\{111\}$ dislocations are in the range $10^{-5} - 10^{-4}\mu$, with considerable measurement uncertainties [32, 14]. Atomistic studies carried out in bcc Fe reveal a slightly higher Peierls stress ($\approx 3 \times 10^{-4}\mu$) for $\langle 111\rangle\{110\}$ edge dislocations [33], and a significantly higher value ($\approx 10^{-2}\mu$) for $\{112\}\langle 111\rangle$ edge dislocations [34, 35]. The Peierls stress, however, is significantly higher ($\approx 10^{-2}\mu$) for $\langle 111\rangle$ screw dislocations in bcc metals, due to their aforementioned non-planar core structure. At 0 K, dislocations cannot glide under shear stresses lower than the Peierls stress. As temperature increases, however, dislocations can move under a stress lower than $\tau_p$ by thermally-assisted nucleation of kink-pairs and their subsequent migration, because they experience much lower lattice resistance to their lateral motion [36]. At a given temperature, the Gibbs free energy ($\Delta G_{kp}$) for kink-pair nucleation is reduced by increasing the applied stress. Corresponding to the stress at which $\Delta G_{kp}$ vanishes, the dislocation appears to move as a whole again, and its velocity is controlled by dissipative mechanisms alone. Therefore, the condition of vanishing $\Delta G_{kp}$ separates two regimes of dislocation mobility, as will be discussed below.

### 2.1. Dislocation velocity controlled by energy dissipation

Let us first consider an infinite straight dislocation subjected to a sufficiently high stress to move uniformly over the Peierls barrier as a whole, i.e. without kinks. The dislocation experiences a friction force as a consequence of energy dissipation caused by several internal mechanisms [37]. Among these, a decisive role is played by dissipative processes in the phonon subsystem of the crystal. In metals, dissipation processes related to the electron subsystem are also present, but typically phonon dissipation dominates, except at low temperatures where the phonon density is limited [38]. Lothe [39] carried out the first comprehensive analysis of the different dissipation mechanisms related to the phonon subsystem. Despite adopting different classification criteria, subsequent work [40, 37, 38] remained within the framework of Lothe's analysis. In the picture that emerged, a non-relativistic dislocation moving with uniform velocity $v$ is described to experience a drag force per unit length

$$f_{\text{drag}} = -B(T)v\,, \qquad (1)$$

where $B(T)$ is a drag coefficient dependent on temperature $T$ and independent of $v$. In this case, three dissipation mechanisms contribute to the drag coefficient, namely thermoelastic dissipation[1], phonon viscosity[2], and phonon scattering[3]. Brailsford [43] developed a unified formalism to describe these different dissipation mechanisms. Based on the unified theory of phonon dissipation, it was concluded that scattering dominates over other mechanisms, and that the corresponding drag coefficients scales as $(T/\Theta_D)^5$ at low temperatures ($T/\Theta_D \ll 1$, with $\Theta_D$ being the Debye temperature), and linearly with temperature otherwise. The linear dependence of the drag coefficient on temperature is in agreement with the original theory of phonon scattering by a moving dislocation developed by Leibfried [44], according to which the drag coefficient is

$$B(T) = \frac{3k_B z T}{10 b^2 c_s}\,, \qquad (2)$$

with $k_B$ being the Boltzmann constant, $z$ the number of atoms in the unit cell, $b$ the magnitude of the Burgers vector, and $c_s$ the shear wave speed. Drag coefficients proportional to temperature have also been consistently found in molecular dynamics (MD) simulations of extended highly mobile dislocations [45, 46, 47].

Temperature-independent contributions to the drag coefficients have also been predicted theoretically. In a unified quantum framework of phonon dissipation, Al'Shitz [48] found a constant contribution to the drag coefficient due to "slow" phonons. Being phonon viscosity dependent on the phonon density $D(\omega)$, such contribution arises considering the non-linear phonon dispersion curve, where $D(\omega)$ increases when the group velocity $d\omega/dk$ is small, such as at the Brillouin zone boundaries [see also 49]. Electron scattering also gives a temperature independent term, [e.g. 50, ch. 7], though this is known to be at least an order of magnitude smaller than phonon scattering contributions above 10% of the Debye temperature. Furthermore, Marian and Caro [51] have seen that in fcc crystals the scattering regime is bounded by phase and group wave velocities along the slip direction, which govern the frequency range of modes susceptible of interacting with the dislocation core.

In a challenge to phonon scattering theories, Swinburne et al [52] recently showed that thermal vibrations around a defect core cannot be described in terms of phonon modes. As a consequence, phonon momentum is not conserved, leading to defect dissipation mechanisms that cannot be described by existing phonon scattering theories. The theory predicts a leading temperature independent contribution to the drag coefficient in terms of the crystal Hessian, in quantitative agreement with molecular dynamics simulations, along with a temperature dependent component that reduces to the dominant term (due to anharmonic strain) of phonon scattering theory in a continuum limit [53]. The temperature independent term can dominate for nanoscale defects in bcc metals, where the temperature dependent term is very small due to their relatively light lattice deformation. In particular, the theory explains the temperature-independence of the drag coefficient of dislocation kinks reported by several authors [54, 55, 56]. In light of these recent developments, in this work we shall consider drag coefficients of the form

$$B(T) = B_0 + B_1 T \qquad (3)$$

for non-relativistic dislocations. In this case, being the drag coefficient independent of the dislocation velocity, force equilibrium for the dislocation yields the mobility law

$$v = \frac{\tau b}{B(T)}, \qquad (4)$$

where $\tau$ is the resolved shear stress and $\tau b$ is the glide component of the Peach-Koehler force.

---

[1] Thermoelastic dissipation is typically negligible in metals due to their high thermal conductivity [39].

[2] The original theory of phonon viscosity is due to Mason [41].

[3] A dislocation scatters phonons by two mechanisms [42]. First, the finite strain induced near the core triggers the non-linear character of the interatomic force law, and, for edge dislocations, it alters the local density (*phonon wind*). Second, the dislocation oscillates under an impinging sound wave, and in doing so it re-radiates energy in the form of cylindrical waves (*flutter mechanism*).

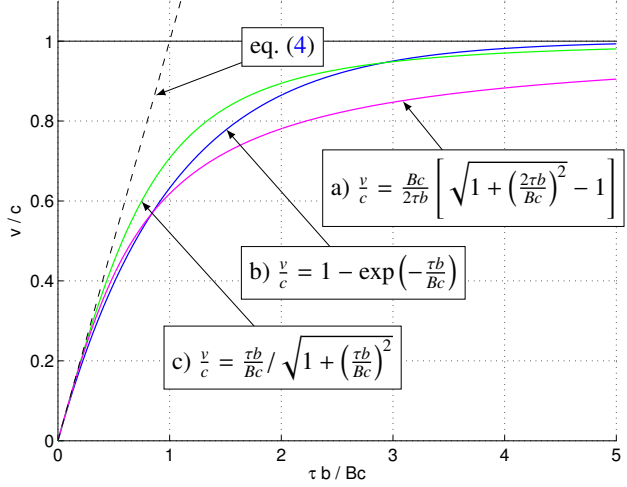

Figure 2: Comparison of parameter-free corrections of eq. (4) for high-speed dislocations. Proposed laws are linear at small stress and tend to an asymptotic value $c$ with increasing stress: a) Taylor et al [59] (b) Gilman [60] (c) Gillis et al [61]: the high velocity drag coefficient is $B = B_0/\sqrt{1-(v/c)^2}$ [see also 62, 63].

The theory considered so far is based on the assumption that dislocation motion is non-relativistic. As the dislocation approaches the speed of sound, however, the mobility law (4) is expected to change because both relativistic effects and dissipation mechanisms active at high speed come into play. In this regime, physical-based theories of drag and mobility become complicated, and they yield ground to empirical laws describing the non-linear dependence of velocity on stress [e.g. 57, 58, 51]. Limiting the attention to subsonic motion, we briefly mention that minimal (i.e. parameter-free) high-speed modifications of (4) have been proposed based on two criteria: (a) that (4) is recovered in the low stress limit, and (b) that the velocity is asymptotically capped by a limiting value $c$. Several laws fulfilling such requirements are compared in fig. 2.

### 2.2. Dislocation velocity controlled by kink-pair nucleation and migration

Let us now consider a straight dislocation line lying in a Peierls valley, subject to a stress lower than the Peierls stress. In this case, the dislocation velocity is determined by the mechanism of kink-pair nucleation and migration.

The first step in obtaining an expression for the dislocation velocity is to model the lateral velocity of kinks[4]. The kink height is $h$, while the secondary Peierls energy barrier is $W_m$ with periodicity $a$. If $W_m$ is low, as in metals, the kink velocity $v_k$ will be determined by phonon drag according to eq. (4), with positive and negative kinks moving in opposite directions. On the other hand, if the secondary Peierls barrier is high, as in covalent materials, kink motion can be treated by the kink-diffusion theory of Lothe and Hirth [64] [see also 50, 65, 66, 67, 68, 14]. In both cases, the kink velocity can be expressed in the form

$$v_k = \frac{\tau b}{B_k(T)}, \qquad (5)$$

with appropriate definition[5] of the effective drag coefficient $B_k(T)$.

When controlled by the mechanism of kink-pair nucleation and migration, the dislocation velocity is determined by the interplay between the kink nucleation rate and the kink migration rate. Because the lateral kink velocity increases with increasing shear stress, as per (5), kink migration is the factor controlling mobility at low stress. Based on this consideration, two limiting regimes are identified [50], as discussed next.

At low stress, kink nucleation time is negligible compared to the kink migration time. The dislocation advances a distance $h$ in the time required for kinks of positive kink-pairs to travel half the distance $2/c_{kp}$ between positive kink-pairs and annihilate with opposite kinks from other pairs, while negative kink-pairs collapse under the applied stress. Therefore, the dislocation velocity is

$$v = h c_{kp} v_k. \qquad (6)$$

---

[4]Because screw dislocations in bcc metals are not orthogonal to planes of crystal symmetry, positive and negative kinks have in general different properties. Here $v_k$ refers to the average velocity of positive and negative kinks.

[5] For the case diffusion-controlled kink velocity, the derivation of (5) goes as follows. If $a$ is the width of the secondary Peierls valleys, the kink drift velocity is $v_k = a(f^+ - f^-)$, where $f^+$ and $f^-$ are the frequencies of forward and backward jumps, respectively. Because kink motion is thermally activated, in the absence of an applied stress both frequencies are equal to $f = \nu_D \exp(-(W_m - TS_m)/k_B T)$, where $\nu_D$ is the Debye frequency. When a stress $\tau$ is applied, diffusion becomes biased and the jump frequencies are:

$$f^\pm = \nu_D \exp\left(-\frac{W_m - TS_m \mp \tau bha/2}{k_B T}\right)$$

In this case, kinks acquire the net drift velocity

$$v_k = 2a\nu_D \exp\left(-\frac{W_m - TS_m}{k_B T}\right) \sinh\left(\frac{\tau bha}{2k_B T}\right)$$

Under the condition $\tau bha/2 \ll k_B T$, the hyperbolic sine can be linearized and the kink velocity takes the form (5) by letting $B_k(T) = \frac{k_B T}{ha^2 \nu_D} \exp(W_m - TS_m/k_B T)$.

Moreover, and because of the low stress, kink-pair concentration $c_{kp}$ does not differ much from the equilibrium concentration $c_k^0 = 2/a \exp(-\Delta G_k/k_BT)$ of positive and negative kinks, and the free enthalpy of kink nucleation $\Delta G_k$ is roughly half that of kink-pair nucleation $\Delta G_{kp}$. Thus, the dislocation velocity turns out to be:

$$v = \frac{2h}{a} v_k \exp\left(-\frac{\Delta G_{kp}}{2k_BT}\right). \qquad (7)$$

On the other hand, at high stress the rate of kink-pair nucleation is the limiting factor. Letting $J_{kp}$ be the net probability of kink-pair nucleation per unit length and time, the dislocation advances a distance $h$ in the time interval $1/J_{kp}X$, where $X$ is the average length available for kink-pair nucleation. Accordingly, the velocity of the dislocation becomes:

$$v = hJ_{kp}X. \qquad (8)$$

Different models have been proposed to estimate the kink-pair nucleation rate $J_{kp}$. In the kink-diffusion theory of Hirth and Lothe [50], the kink-pair nucleation rate equals the total diffusional flux, which is:

$$J_{kp} = \frac{v_k}{a^2} \exp\left(-\frac{\Delta G_{kp}}{k_BT}\right). \qquad (9)$$

The length $X$ is determined by two possible conditions. For sufficiently long dislocations, kinks annihilate when they collide with opposite kinks from neighboring kink-pairs, so that $X$ must be the average distance $\lambda = 1/c_{kp}$ between kink-pairs. For shorter dislocation segments, $X$ is the length $L$ of the segment. The geometric average of these two quantities is a suitable measure for $X$:

$$X = \frac{\lambda L}{\lambda + L}. \qquad (10)$$

The quantity $\lambda$ is found from the condition of steady-state motion of the dislocation, which requires that the migration time between annihilation events be equal to the nucleation time over the average length of the growing kink-pair. This reads:

$$\frac{\lambda/2}{v_k} = \frac{1}{J_{kp}\lambda/2}. \qquad (11)$$

Solving for $\lambda$ yields

$$\lambda = 2\sqrt{\frac{v_k}{J_{kp}}} = 2a \exp\left(\frac{\Delta G_{kp}}{2k_BT}\right). \qquad (12)$$

Finally, substituting eq. (10) and eq. (9) into eq. (8), the dislocation velocity is found to be

$$\begin{aligned} v &= \frac{2hL}{\lambda + L}\sqrt{J_{kp}v_k} \\ &= \frac{L}{2a\exp\left(\frac{\Delta G_{kp}}{2k_BT}\right) + L} \frac{2h}{a} v_k \exp\left(-\frac{\Delta G_{kp}}{2k_BT}\right). \end{aligned} \qquad (13)$$

In covalent crystals, the kink velocity is low because it is controlled by diffusion, and therefore it is possible to consider the *kink-collision* regime $\lambda \ll L$, where the dislocation velocity becomes length independent. In this case, eq. (13) reduces to (7), extending its validity over the entire stress range. On the other hand, in metals, the higher kink velocity controlled by phonon-drag corresponds to the *length-dependent* regime, where $\lambda \gg L$. In this case, the dislocation velocity is:

$$v = \frac{hL}{a^2} v_k \exp\left(-\frac{\Delta G_{kp}}{k_BT}\right). \qquad (14)$$

The proportionality between dislocation velocity and length has been confirmed experimentally in bcc metals [69, 70], and in atomistically-informed Kinetic Monte Carlo (KMC) simulations [55]. The length dependence is expected to saturate for sufficiently long dislocations. This, and the fact that the pre-exponential term scales approximately linearly with stress, was confirmed by Deo et al [71] in KMC simulations of screw dislocations velocity in bcc Ta.

In the model of Dorn and Rajnak [72], the rate of kink-pair nucleation follows an Arrhenius-type law

$$J_{kp} = \frac{1}{w} v_D \exp\left(-\frac{\Delta G_{kp}}{k_BT}\right), \qquad (15)$$

where $v_D$ is the Debye frequency, and $w$ is the kink-width. Using eq. (15) into (8) with $X = L$ yields a result similar to eq. (14), but with a pre-exponential term independent of $v_k$ (hence of stress) and where the ratio $L/w$ is the total number of kink-pair "nucleation sites":

$$v = \frac{L}{w} h v_D \exp\left(-\frac{\Delta G_{kp}}{k_BT}\right). \qquad (16)$$

The parameter $w$ is in general dependent on stress, but (16) has been used in several DD models of screw dislocation mobility with a constant pre-exponential term [20, 23, 25, 73]. In this case, (16) predicts a net dislocation velocity at zero stress. This artifact arises because (16) neglects the probability of backward jumps. In order to remedy this effect, the exponential has sometimes been replaced by a hyperbolic sine [e.g. 21]. However, the physical justification of such function has been questioned in the literature [74, 75].

Variations of the model of Dorn and Rajnak [72] have also been considered. For example, Gilbert et al [76] used eq. (15) in conjunction with the condition (11) to eliminate $L$, and found the result

$$v = h\sqrt{\frac{2v_k v_D}{w}} \exp\left(-\frac{\Delta G_{kp}}{2k_BT}\right). \qquad (17)$$

Using eq. (15), Cereceda et al [56] defined the dislocation velocity dividing the distance $h$ by the sum of the kink-pair nucleation and migration times:

$$\begin{aligned} v &= \frac{h}{\frac{1}{J_{K_p}(L-w)} + \frac{L-w}{2v_k}} \\ &= \frac{2\tau b h(L-w) v_D \exp\left(-\frac{\Delta G_{kp}}{k_BT}\right)}{2\tau bw + (L-w)^2 B_k v_D \exp\left(-\frac{\Delta G_{kp}}{k_BT}\right)}, \end{aligned} \qquad (18)$$

which at low stresses and/or temperatures reduces to the diffusive form (16), while at high stresses/temperatures the velocity is dominated by the phonon drag term, akin to eq. (4).

### 2.3. The free energy of kink-pair formation

The dislocation velocities derived in the previous section involve the Gibbs free energy of kink-pair formation:

$$\Delta G_{kp} = \Delta H_{kp} - T\Delta S_{kp} \tag{19}$$

where $\Delta H_{kp}$ is the formation enthalpy, and $\Delta S_{kp}$ is the corresponding entropy. At low stress, positive and negative kinks of a pair are well-separated and, according to Seeger and Schiller [18], the enthalpy of kink-pair formation includes twice the formation energy of one kink, the elastic interaction energy of the two kinks, and the work of the applied stress. If the two kinks are a distance $x$ apart, the enthalpy reads:

$$\Delta H_{kp}(x) = 2U_k - \frac{\mu b^2 h^2}{8\pi x} - \tau b h x. \tag{20}$$

The equilibrium configuration of the kink-pair is reached for a separation distance $x = (\mu bh/8\pi\tau)^{1/2}$, which, when substituted back into the previous equation yields an equilibrium enthalpy

$$\Delta H_{kp} = 2U_k \left[1 - \left(\frac{\tau}{\tau_I}\right)^{1/2}\right], \tag{21}$$

where $\tau_I = 8\pi U_k^2/\mu b^3 h^3$ is a cut-off stress above which $\Delta H_{kp}$ vanishes.

The previous result is valid at low stress, when the separation distance between kinks is much larger than the kink width. For higher stresses Dorn and Rajnak [72] proposed a model where $\Delta H_{kp}$ is computed by minimization of an energy functional of the kink-pair shape. For an anti-parabolic Peierls potential, the following result holds [2]:

$$\Delta H_{kp} = 2U_k \left[1 - \left(\frac{\tau}{\tau_{II}}\right)\right]^2. \tag{22}$$

In practice, different forms of kink-pair activation enthalpy can be fitted by the phenomenological law introduced by Kocks at al [77]:

$$\Delta H_{kp} = \Delta H_0 \left[1 - \Theta(\sigma)^p\right]^q, \tag{23}$$

where $0 \leq p \leq 1$ and $1 \leq q \leq 2$ are fitting parameters, and $\Theta = \tau/\tau_P$ is the ratio between the resolved shear stress and the Peierls stress.

The entropy $\Delta S_{kp}$ is more difficult to model. Nevertheless, its temperature and stress dependence are predicted to be mild (logarithmic) [78]. Typically, the term $\exp(\Delta S_{kp}/k_B)$ is absorbed into the pre-exponential of the dislocation velocity expression as a constant multiplicative contribution. However, it is convenient to write a constant entropic term as

$$\Delta S_{kp} = \frac{\Delta H_0}{T_0}, \tag{24}$$

so that the free enthalpy becomes

$$\Delta G_{kp}(\sigma, T) = \Delta H_0 \left\{ [1 - \Theta(\sigma)^p]^q - \frac{T}{T_0} \right\}. \tag{25}$$

In this form, the fitting constant $T_0$ can be interpreted an athermal transition temperature, above which the energy barrier for kink-pair nucleation is guaranteed to vanish independent of stress. Note that the function $\Delta G_{kp}(\sigma, T)$ defines the domain of active kink-pair mechanism ($\Delta G_{kp} > 0$), and the condition $\Delta G_{kp} = 0$ marks the transition to the regime of dislocation velocity controlled by phonon-drag alone.

### 2.4. Non-Schmid effects

The main drawback of the definition of the stress ratio $\Theta$ given after (23) is that it enforces Schmid's law and therefore it fails to capture non-Schmid effects characteristic of bcc plasticity. However, recent work has shown that the Kocks model can be generalized to capture non-Schmid effects with an appropriate definition of the stress ratio $\Theta$. According to Hale et al [80], the definition of the stress ratio $\Theta$ which best extends the validity of Kocks law in the presence of non-Schmid stresses is

$$\Theta(\sigma) = \frac{\tau_{\text{mrss}}}{\tau_{\text{crss}}(\sigma)}, \tag{26}$$

where $\tau_{\text{mrss}}$ is the *maximum resolved shear stress*[6], and its critical value, i.e. the value corresponding to rigid motion of the dislocation over the Peierls barrier, is the *critical resolved shear stress* $\tau_{\text{crss}}$. Note that, in contrast with Schmid's law, both $\tau_{\text{mrss}}$ and $\tau_{\text{crss}}$ may be relative to non-crystallographic planes, and that the latter represents a stress-dependent critical value.

We now discuss the functional dependance of $\tau_{\text{crss}}$ on the state of stress. According to Gröger et al [81, 79], the condition of initiation of glide at 0 K of a [111]($\bar{1}$01) screw dislocation can be described by the following expression[7]:

$$\tau_{(\bar{1}01)} + a_1\tau_{(0\bar{1}1)} + a_2\tau'_{(\bar{1}01)} + a_3\tau'_{(0\bar{1}1)} = a_0\tau_p. \tag{27}$$

The meaning of the shear stresses appearing in eq. (27) is illustrated in fig. 3. These are:

$$\tau_{(\bar{1}01)} = \sigma : \mathbf{m}^\alpha \otimes \mathbf{n}^\alpha \tag{28a}$$

$$\tau_{(0\bar{1}1)} = \sigma : \mathbf{m}^\alpha \otimes \mathbf{n}_1^\alpha \tag{28b}$$

$$\tau'_{(\bar{1}01)} = \sigma : (\mathbf{n}^\alpha \times \mathbf{m}^\alpha) \otimes \mathbf{n}^\alpha \tag{28c}$$

$$\tau'_{(0\bar{1}1)} = \sigma : (\mathbf{n}_1^\alpha \times \mathbf{m}^\alpha) \otimes \mathbf{n}_1^\alpha \tag{28d}$$

where $\tau_{(\bar{1}01)}$ is the stress component parallel to the Burgers vector on the glide plane (glide stress); $\tau_{(0\bar{1}1)}$ is the shear stress parallel to the Burgers vector on the $(0\bar{1}1)$ plane at 60° to the glide plane, and it is responsible for the T/AT asymmetry of the flow stress. The stress components $\tau'_{(\bar{1}01)}$ and $\tau'_{(0\bar{1}1)}$ are shear stresses orthogonal to the Burgers vector, in the $(\bar{1}01)$ and $(0\bar{1}1)$ planes, respectively, and they are responsible for the T/C asymmetry of the flow stress. The constants $a_0, \ldots a_3$ are fitting parameters obtained from the atomistic calculations.

---

[6]$\tau_{\text{mrss}}$ is the stress resolved on the maximum resolved shear stress plane (mrssp), i.e. the (potentially non-crystallographic) plane with highest Schmid factor in the zone of the slip direction.

[7]For alternative yield criteria see [82, 83, 84] and references therein.

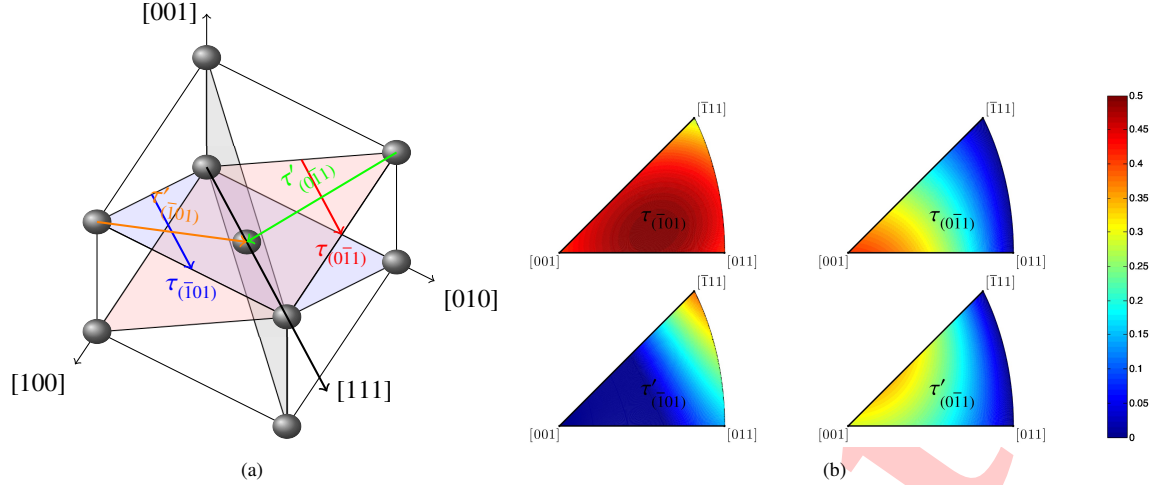

Figure 3: Stress projections responsible for non-Schmid effects according to [79]. (a) Schematic representation of the stress components $\tau_{(\bar{1}01)}$, $\tau_{(0\bar{1}1)}$, $\tau'_{(\bar{1}01)}$, and $\tau'_{(0\bar{1}1)}$ used in the model of to account for non-schmid effects. (b) Schmid factors for each stress component in case of a uniaxial load in the standard triangle.

Note that only two of the four stress projections used in (27) depend on $\tau_{\text{mrssp}}$, and in particular:

$$\tau_{(\bar{1}01)} = \tau_{\text{mrss}} \cos(\chi) \qquad (29)$$

$$\tau_{(0\bar{1}1)} = \tau_{\text{mrss}} \cos(\chi + \pi/3) \qquad (30)$$

Together with the condition $\tau_{\text{mrss}} = \tau_{\text{crss}}$ reached at the onset of slip, substituting the previous equations in (27) yields the following expression for $\tau_{\text{crss}}$

$$\tau_{\text{crss}}(\boldsymbol{\sigma}) = \frac{a_0 \tau_p - a_2 \tau'_{(\bar{1}01)} - a_3 \tau'_{(0\bar{1}1)}}{\cos(\chi) + a_1 \cos(\chi + \pi/3)} \qquad (31)$$

and the corresponding stress ratio $\Theta$:

$$\Theta = \frac{\tau_{(\bar{1}01)} + a_1 \tau_{(0\bar{1}1)}}{a_0 \tau_p - a_2 \tau'_{(\bar{1}01)} - a_3 \tau'_{(0\bar{1}1)}} \qquad (32)$$

## 3. A phenomenological mobility law for bcc crystals

Guided by the theory presented in the previous section, our next goal is to formulate a mobility law for bcc metals to be used in discrete and continuum DD simulations. Tungsten (W), is our model material of choice, and some of its relevant physical properties are listed in table 1. We limit our formulation to the *glide* component of dislocation motion. Dislocation *climb*, being controlled by diffusion of point defects into the dislocation core, is beyond the current scope of this work.

In developing the mobility law, we picture a mesoscopic description of dislocation lines in their glide planes where kinks are not explicitly tracked, and where the line orientation changes smoothly and continuously along the dislocation line. Because the only physical component of line motion is that normal to itself, we express the glide velocity of a dislocation as

$$\boldsymbol{v} = v\hat{\boldsymbol{v}} \qquad (33)$$

where $\hat{\boldsymbol{v}} = \hat{\boldsymbol{n}} \times \hat{\boldsymbol{\xi}}$ is a unit vector lying in the slip plane with unit normal $\hat{\boldsymbol{n}}$, and orthogonal to the dislocation unit tangent $\hat{\boldsymbol{\xi}}$. The vector direction being known, we shall focus our attention on the magnitude $v$, which in general is assumed to be a function of the local stress tensor $\boldsymbol{\sigma}$ and temperature $T$, and of a series of parameters specific to the slip-system and to the dislocation character (screw, edge, or mixed). The latter is defined by the angle $\varphi$ between the Burgers vector and the line direction $\boldsymbol{\xi}$. Recent atomistic simulations [87] suggest that the transition in Peierls stress between screw and edge character is non-smooth, with several discrete peaks as a function of $\varphi$ corresponding to low-index orientations. For simplicity, we only here consider interpolation between pure screw and edge dislocation velocities, although more interpolation points as a function of $\varphi$ could in principle be used. Therefore the general form of the proposed mobility law is:

$$v(\boldsymbol{\sigma}, T, \varphi) = v_s(\boldsymbol{\sigma}, T)[1 - w(\varphi)] + v_e(\boldsymbol{\sigma}, T)w(\varphi). \qquad (34)$$

| melting temperature | $T_m$ | 3695 [K] |
| lattice constant | $a$ | 3.16 [Å] |
| Burgers vector | $b$ | $a\sqrt{3}/2$ |
| Peierls stress | $\tau_p$ | 2.03 [GPa] |
| kink height | $h$ | $a\sqrt{2/3}$ |
| kink width | $w$ | 25a |

Table 1: Relevant physical properties of bcc W and $\langle 111 \rangle \{110\}$ slip. Peierls stress and kink width are taken from [86].

In (34), $w(\varphi)$ is an interpolation function with properties $w(0) = 0$, and $w(\pi/2) = 1$. Our numerical results are obtained with the choice $w(\varphi) = \sin^2 \varphi$. Edge and screw velocities are assumed to have the same general functional form but they are parametrized by different coefficients. These parameters are in general slip-system specific. This allows to account for kink-controlled edge dislocation velocity on slip systems other than $\langle 111 \rangle \{110\}$ [34, 35].

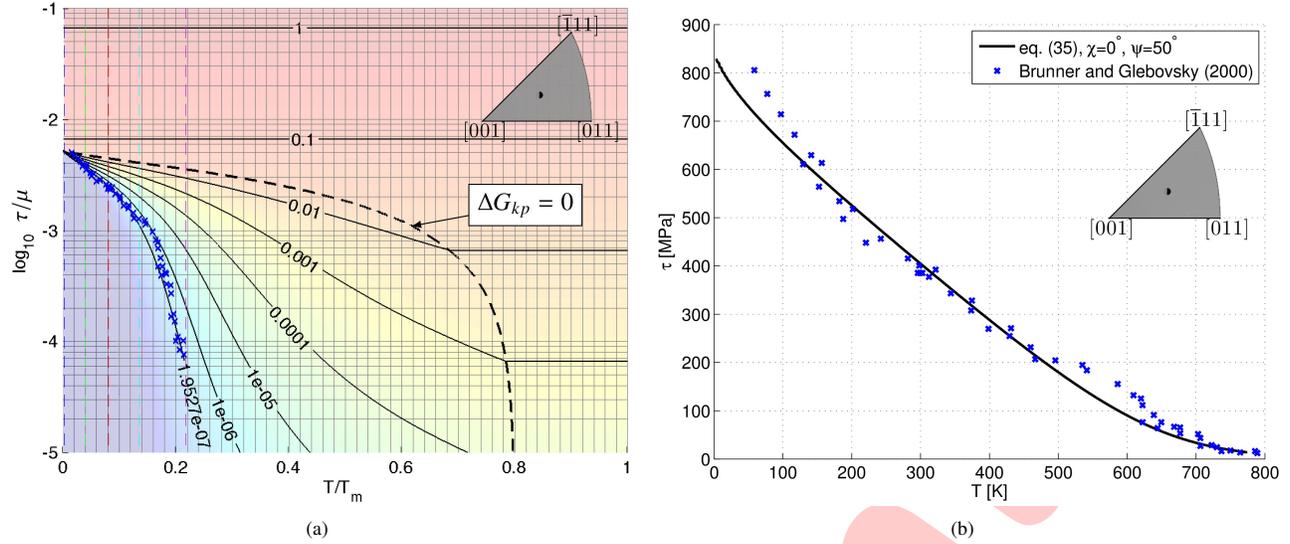

Figure 4: (a) Map of screw dislocation velocity contours as a function of normalized temperature and normalized resolved shear stress, obtained from eq. (35) for bcc W. Black lines representing $v/c_s$ isovalues for a $[111](\bar{1}01)$ screw dislocation subjected to a uniaxial tensile load at $50°$ from the slip direction and $\chi = 0$, as indicated in the upper-right inset of the stereographic triangle. This loading mode corresponds to the experimental conditions of [85]. The isovalue $v/c \approx 1.95 \times 10^{-7}$ corresponds to the velocity computed from Orowan's equation $\dot{\gamma} = \rho b v$, where the values $\dot{\gamma} = 8.5 \times 10^{-4}\ [s^{-1}]$, $\rho = 5.5 \times 10^9\ [m^{-2}]$, and $b = 2.736 \times 10^{-10}\ [m]$ are taken from [85]. For comparison, velocity values derived from the flow stress data of [85] are shown by crosses. (b) Calculated flow stress corresponding to $\dot{\gamma} = 8.5 \times 10^{-4}\ [s^{-1}]$ as a function of temperature compared to the experimental data of [85].

A criterion guiding the formulation of the mobility law is that it should allow for a smooth transition between kink-dominated and phonon dominated regimes as a function of stress and temperature. Because of short-range dislocation-dislocation interactions, stress can vary rapidly along dislocation lines, and both regimes may be encountered within the simulation domain. With this requirement, the general form of the dislocation velocity (edge or screw) is implemented as:

$$v(\boldsymbol{\sigma}, T) = \begin{cases} \frac{\tau b}{B(\boldsymbol{\sigma},T)} \exp\left(-\frac{\Delta G_{kp}(\boldsymbol{\sigma},T)}{2k_B T}\right) & \text{if } \Delta G_{kp}(\boldsymbol{\sigma},T) > 0 \\ \frac{\tau b}{B(\boldsymbol{\sigma},T)} & \text{if } \Delta G_{kp}(\boldsymbol{\sigma},T) \leq 0 \end{cases} \quad (35)$$

In Eq. (35), the free energy of kink-pair nucleation $\Delta G_{kp}$ is defined according to (25). The condition $\Delta G_{kp} > 0$ defines the regime of active kink-pair mechanism. In this regime, we let $B(\boldsymbol{\sigma}, T) = B_k$, so that (35) coincides with the kink-diffusion model (13). Among the kink-pair models mentioned in the previous section, the kink-diffusion model (13) is found to provide the best fit with experimental flow-stress data, as shown in fig. 4(b). On the other hand, when $\Delta G_{kp} = 0$, (35) takes the general form (4) describing the velocity controlled by phonon drag provided that $B(\boldsymbol{\sigma}, T) = B_0 + B_1 T$. Therefore, both regimes of dislocation velocity are captured by (35) with the following definition of the drag coefficient:

$$B(\boldsymbol{\sigma}, T) = \begin{cases} \frac{a\left[2a\exp\left(\frac{\Delta G_{kp}(\boldsymbol{\sigma},T)}{2k_B T}\right) + L\right]}{2hL} B_k & \text{if } \Delta G_{kp}(\boldsymbol{\sigma},T) > 0 \\ B_0 + B_1 T & \text{if } \Delta G_{kp}(\boldsymbol{\sigma},T) \leq 0 \end{cases} \quad (36)$$

A smooth transition between these two functions is obtained with a sigmoidal function centered around $\Delta G_{kp} = 0$.

Next we describe the procedure used to obtain the fitting parameters involved in (35), for both edge and screw $\langle 111 \rangle \{110\}$ dislocations in W. All atomistic data are obtained using the EAM potential of Marinica et al [88]. This potential was found to be particularly suitable to describe both static and dynamic properties of screw dislocations in W [86], based on the comparison of five different interatomic potentials.

### 3.1. Fitting procedure for $\langle 111 \rangle \{110\}$ edge dislocations

For edge dislocations, we take $\Delta H_0 \approx 0$, so that all other parameters related to the kink-pair model are not required, and the only values needed are the drag coefficients $B_0$ and $B_1$. A summary of the MD simulation procedure is given next.

A $1/2\langle 111 \rangle (10\bar{1})$ edge dislocation dipole was formed in a supercell of dimensions $60\|111\|a \times 60\|10\bar{1}\|a \times \|1\bar{2}1\|a$, containing 21960 atoms. The dislocation slip planes were separated by $30\|10\bar{1}\|a$. At T = 200, 300, 400, and 500 K, the system was first equilibrated over 0.1 ns before being run under microcanonical conditions for 1 ns, under no applied stress. The dislocation core positions were extracted every 200 fs via a time averaged energy filter as described in [54]. The dislocation dipole was observed to undergo free diffusion, with a center of mass diffusion constant $D_{com} = D/2$, where D is the diffusivity of an individual dislocation. Using the Einstein relation

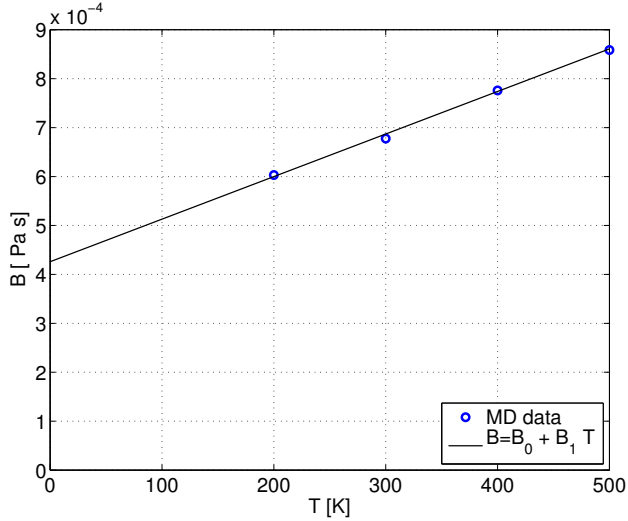

Figure 5: Phonon drag coefficient $B = B_0 + B_1 T$ for $1/2\langle 111\rangle(10\bar{1})$ edge dislocations in W. The coefficients $B_0 = 4.26 \times 10^{-4}$[Pa s] and $B_1 = 0.87 \times 10^{-6}$[Pa s/K] have been extracted from MD simulations using the EAM potential of Marinica et al [88].

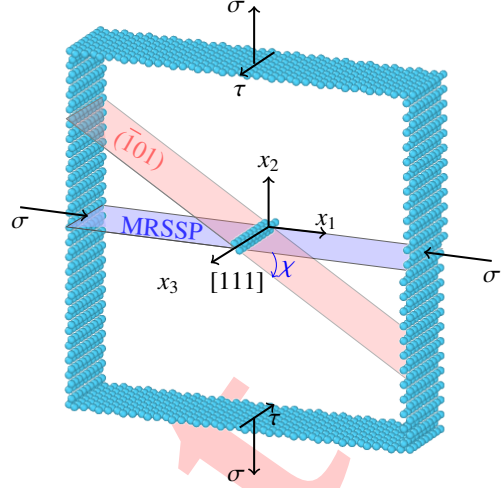

Figure 6: Schematic representation of the NEB simulation setup used to determine $\tau_{\text{crss}}$ as a function of the angle $\chi$ and the stress $\sigma$.

$D = kT/(LB)$, where $L = \|121\|a$, the drag parameter B was extracted at each temperature. This diffusion-based, zero-stress simulation method to measure dislocation drag has been shown to be in very good agreement with drift-based, stress-driven simulations [89], a demonstration of the fluctuation-dissipation relation as applied to dislocations. Simulation results are shown in fig. 5, and the values of the fitting parameters $B_0$ and $B_1$ are reported in table 2.

### 3.2. Fitting procedure for $\langle 111\rangle\{110\}$ screw dislocations

For screw dislocations, a temperature independent kink drag coefficient $B_k = 8.3 \times 10^{-5}$ [Pa s] is used [55]. Based on the Frenkel-Kontorova model [54], this value corresponds to a constant line drag coefficient $B_0 = 9.8 \times 10^{-4}$ [Pa s]. Parameters related to the free energy of kink-pair nucleation are taken from the literature [55]. These are the pre-factor of Kocks law $\Delta H_0 = 1.63$ [eV], and the two exponents $p = 0.86$, and $q = 1.69$.

In order to study the dependence of $\tau_{\text{crss}}(\sigma)$ on the state of stress and characterize the non-Schmidt model (26), we perform a series of nudged elastic band (NEB) calculations using the LAMMPS code ([90, 91]). The simulation boxes contain one single $\langle 111\rangle\{110\}$ screw dislocation per box. With reference to fig. 6, the orientation of the box is such that the $x_3$ axis is along the dislocation line, and the $x_1$ axis is in the MRSSP of the [111] zone. The $(\bar{1}01)$ slip plane is at an angle $\chi$ to the MRSSP. Periodic boundary conditions are applied in the dislocation line direction $x_3$ while non-periodic and shrink-wrapped boundary conditions are applied in the other two directions. Following Gröger et al [81, 79], the stress state applied to the simulation box is:

$$\boldsymbol{\sigma} = \begin{bmatrix} -\sigma & 0 & 0 \\ 0 & \sigma & \tau \\ 0 & \tau & 0 \end{bmatrix}. \tag{37}$$

By varying $\chi$ and $\sigma$, $\tau_{\text{crss}}(\sigma)$ is determined as the value of the applied shear stress $\tau$ that results in a dislocation slip event. Note that, at any $\chi$, the stress component $\sigma$ is the maximum shear stress in the plane orthogonal to the Burgers vector. This quantity is an invariant for the stress components acting in the plane orthogonal the the Burgers. These components couple to the in-plane edge components of the core, and therefore affect the flow stress. The dependency of $\tau_{\text{crss}}$ on $\chi$ is a manifestation of the T/AT asymmetry of the flow stress, while its dependency on $\sigma$ is due to the TC asymmetry. Results of the NEB simulations are shown in fig. 7.

For the stress state (37), the non-Schmid components (28) are:

$$\tau_{(\bar{1}01)} = \tau \cos(\chi) , \tag{38a}$$

$$\tau_{(0\bar{1}1)} = \tau \cos\left(\chi + \frac{\pi}{3}\right) , \tag{38b}$$

$$\tau'_{(\bar{1}01)} = \sigma \sin(2\chi) , \tag{38c}$$

$$\tau'_{(0\bar{1}1)} = \sigma \sin\left(2\chi + \frac{2\pi}{3}\right) . \tag{38d}$$

Using these identities, the parameters $a_0 \ldots a_3$ defining $\tau_{\text{crss}}(\sigma)$ in (31) can be fitted to the NEB dataset. However, in order to guarantee that $\tau_{\text{crss}}$ remains positive for large $\sigma$, and to provide a better fit of the atomistic results, we propose a modified version of (31):

$$\tau_{\text{crss}} = \frac{a_0 \tau_p \, f\left(\frac{a_2 \tau'_{(\bar{1}01)} + a_3 \tau'_{(0\bar{1}1)}}{a_0 \tau_p}\right)}{\cos(\chi) + a_1 \cos(\chi + \pi/3)} , \tag{39}$$

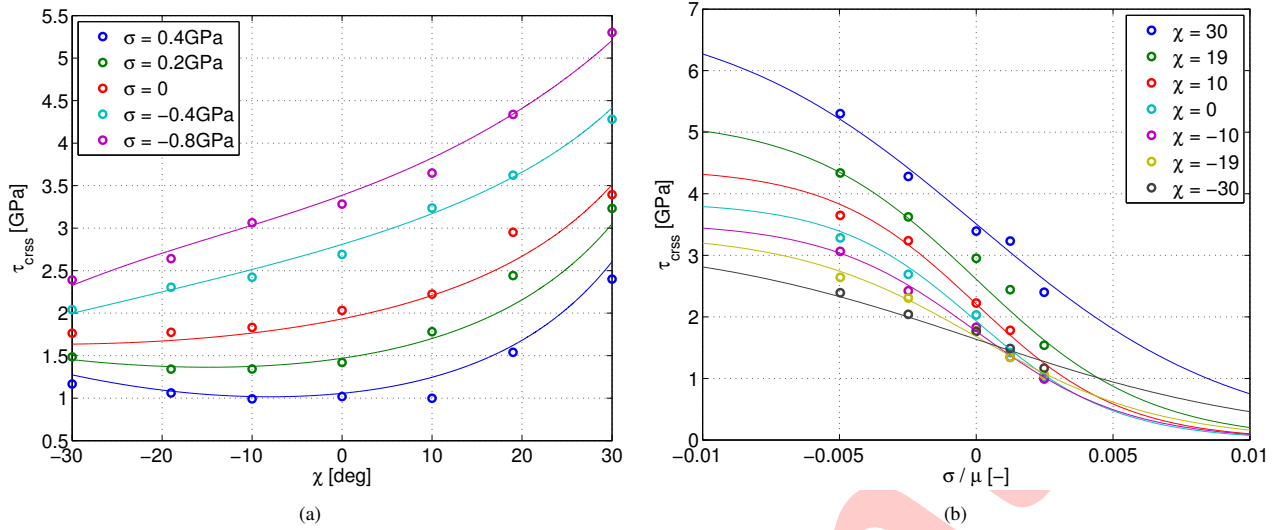

Figure 7: Results of NEB calculations of the critical resolved shear stress $\tau_{\text{crss}}$ and fit by eq. (39), showing its dependence on (a) the orientation of the MRSSP, i.e. the angle $\chi$, and (b) the non glide stresses $\sigma$ in tension and compression.

where $f(x) = 2/(1+e^{2x})$ is a modified sigmoidal function which guarantees that (39) tends to (31) when its argument is small, but remains positive and bounded for all values of its argument. The result of the fitting procedure is shown by solid lines in fig. 7(b), and the value of the fitting constants are reported in table 2. With (39), the stress ratio $\Theta$ used in the mobility law is:

$$\Theta = \frac{\tau_{(\bar{1}01)} + a_1 \tau_{(0\bar{1}1)}}{a_0 \tau_p \, f\left(\frac{a_2 \tau'_{(\bar{1}01)} + a_3 \tau'_{(0\bar{1}1)}}{a_0 \tau_p}\right)} . \quad (40)$$

Finally, with all other parameters determined, the athermal transition temperature $T_0$ is found by the following procedure. We consider single-crystal flow stress data vs temperature of [85], which was obtained for an applied strain rate $\dot{\gamma} = 8.5 \times 10^{-4}\, s^{-1}$ and initial dislocation density $\rho = 5.5 \times 10^9\, [m^{-2}]$. We assume that the flow stress corresponds to the condition that the screw dislocation velocity reaches the value $v = \dot{\gamma}/\rho b$. With this condition, $T_0$ is chosen so that the stress vs. temperature profile of the mobility law (35) best fits the experimental data. Results of the fit are shown in fig. 4, while the value of $T_0$ is reported in table 2.

## 4. Tungsten micropillar deformation

To demonstrate the utility of the developed dislocation mobility law, here we use DD simulations to investigate the influence of temperature and loading orientation on the plastic deformation and corresponding dislocation microstructure in W micro-pillars. The system under investigation is a prismatic pillar with side length of $4000b$ ($\approx 1.1\mu m$) and aspect ratio 2:1. The pillar is populated with an initial density of Frank-Read (FR) sources on random slip systems, and random locations within the pillar. The initial dislocation density is $\approx 10^{13} m^{-2}$.

The DD simulation is coupled with a Finite Element (FE) solution to a superimposed elastic boundary value problem, in which traction corrections are applied on the system boundary to satisfy imposed boundary conditions. The bottom surface of the pillar is fixed, the lateral surfaces are traction-free, and a uniform tensile load is applied to the top surface. The traction increases linearly with time, with rate $\dot{\sigma}/E = 0.415 s^{-1}$, where $E$ is the Young modulus. This loading mode is chosen to highlight the influence of temperature and loading orientation on the flow stress. However, because traction is imposed, the present stress-strain curves don't show the stochastic behavior typically observed in experiments for micro-pillars. For results obtained using an imposed strain-rate, as opposed to a stress-rate, we refer to [92]. Two different temperatures are considered in our DD simulations, namely 300 and 900 K. For each temperature, three crystal orientations are chosen as shown in the inset of fig. 8(a) by the stereographic triangle. The curves plotted in black correspond to the orientation of used in the experiments by Brunner and Glebovsky [85], while other curves are obtained varying the angle $\chi$. Engineering stress-strain curves are obtained for each case. The influence of the sample temperature on the flow stress is clearly observed. Similar to bulk bcc crystals, the lower the temperature, the higher the flow stress. For the same temperature, it can also be observed that the flow stress is higher in the anti-twinning orientation ($\chi > 0$) compared to the twinning orientation ($\chi < 0$), in agreement with fig. 7(a). Note that non-Schmidt effects also decrease with increasing temperature.

|  |  | $\langle 111\rangle\{110\}$ screw | $\langle 111\rangle\{110\}$ edge |
|---|---|---|---|
| $\Delta H_0$ | [eV] | 1.63 | $\approx 0$ |
| $p$ | [-] | 0.86 | - |
| $q$ | [-] | 1.69 | - |
| $T_0$ |  | $0.8 T_m$ | - |
| $\tau_p$ | [GPa] | 2.03 | - |
| $a_0$ | [-] | 1.50 | - |
| $a_1$ | [-] | 1.15 | - |
| $a_2$ | [-] | 1.55 | - |
| $a_3$ | [-] | 2.86 | - |
| $B_k$ | [Pa s] | $8.3 \times 10^{-5}$ | - |
| $B_0$ | [Pa s] | $9.8 \times 10^{-4}$ | $4.26 \times 10^{-4}$ |
| $B_1$ | [Pa s /K] | $\approx 0$ | $0.87 \times 10^{-6}$ |

Table 2: Dislocation mobility parameters for the $\langle 111\rangle\{110\}$ slip system in W.

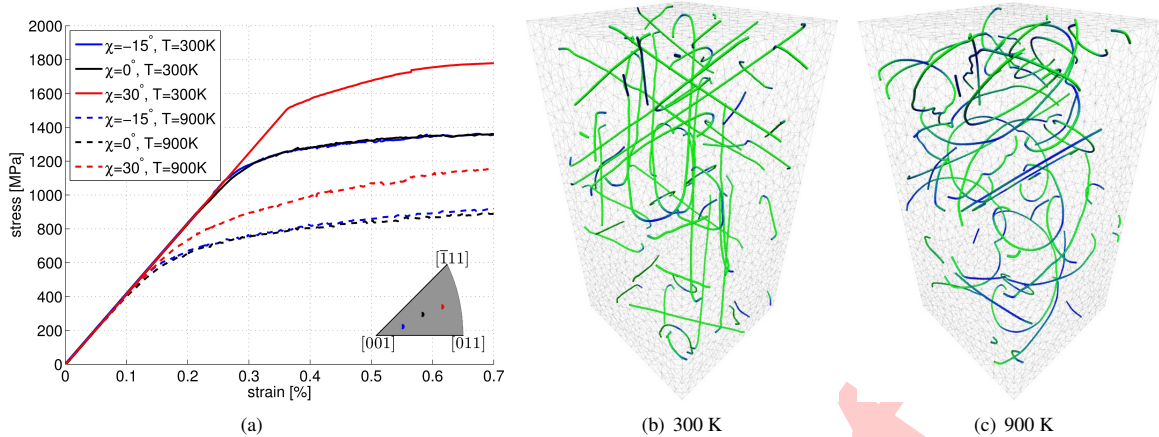

Figure 8: Results of DD simulations of W micro-pillars subject to tensile loading using the mobility law (35). The micro-pillars have side length of $\approx 1.1 \mu m$ and aspect ratio 2:1. (a) Engineering stress-strain curves for two temperatures, 300 and 900 K, and three loading orientations marked in the inset ($\chi = -15, 0, 30°$, $\psi = 50°$). (b) Typical dislocation microstructure configuration at 300 K. Colors represent the angle between the Burgers vector and the local line tangent vector, green for screw character, blue for edge character. (c) Dislocation microstructure at 900 K.

In order to gain insight into the underlying dislocation mechanisms, close examination of the dislocation configuration evolution is carried out. Two typical dislocation configurations are given in fig. 8(b) and fig. 8(c). As expected, at room temperature long screw dislocations can be predominantly observed due to their low mobility. At higher temperature (900 K), dislocation lines become curved, and acquire a mixed screw-edge character. In addition, FR and single arm (SA) sources also exhibit different features, as the temperature is increased. At low temperatures, FR and SA sources operate mainly by bowing out of the mixed part of the dislocation. After it interacts with the free surface, a long screw dislocation is left inside the pillar. This is very similar to experimental observations in Fe [69]. By way of contrast, at high temperatures, SA act similar to half of a Frank-Read source, much like in fcc micro-pillars.

To make further comparison with experimental results, we performed a second set of DD simulations in order to measure the 0.2% yield strength of W micro-pillars of different sizes at room temperature, under both tensile and compressive loads along the $\langle 001 \rangle$ direction. Results are shown in fig. 9, in comparison to experimental data from [93]. For each size, several DD simulations are carried out with different random initial conditions. The simulated yield stress data is in good agreement with the experimental values at all sizes, and it shows the asymmetry of the yield stress in tension and compression.

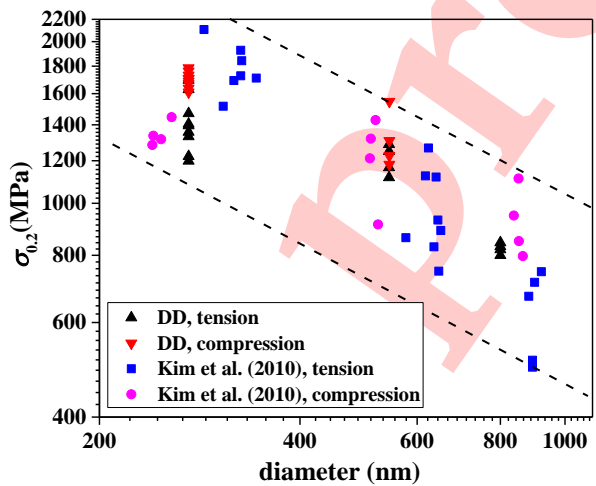

Figure 9: Yield stress of [001] micro-pillars at room temperature vs pillar diameter. Comparison between DD results and experimental values from [93].

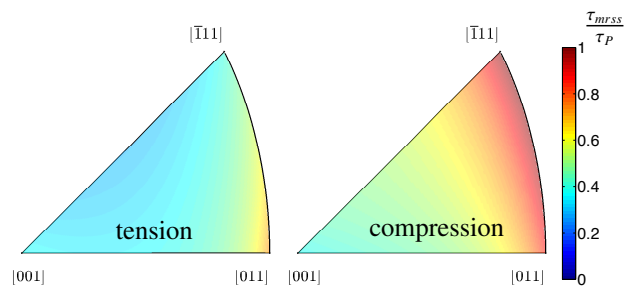

Figure 10: Predicted shear stress resolved on the MRSSP necessary to yield a strain rate of $\approx 10^{-4} \ s^{-1}$ at 77K in W (nominal dislocation density is $\rho = 5.5 \times 10^9 \ m^{-2}$). The $\langle 011 \rangle$ direction is the hardest in tension, while $\langle 111 \rangle$ is the hardest direction in compression, consistently with the behavior of other bcc transition metals [6].

Finally, we consider the influence of loading orientation on the flow stress in tension and compression at low temperature. Duesbery [6] collected values of the the critical shear stress (resolved on the MRSSP) of various bcc metals for the three orientations at the corners of the reference triangle, in both tension and compression. The experimental data corresponds to a temperature of 77 K and a strain rate of $10^{-4}$ $s^{-1}$. The data indicates that the $\langle 011 \rangle$ direction is the hardest in tension, while the $\langle 111 \rangle$ direction is the hardest in compression. A favorable qualitative comparison between this general feature of bcc transition metals and the prediction of the proposed mobility law is shown in fig. 10.

## 5. Summary and Conclusions

In this work, we have formulated a comprehensive phenomenological dislocation mobility law for bcc metals, and have applied the resulting law in large-scale DD simulations of sub micron plasticity in W micro-pillars. In the proposed model, the condition that the free energy of kink-pair nucleation vanish marks the transition between two regimes of dislocation motion, the kink-controlled regime, and the phonon drag regime. Based on existing theoretical framework, the proposed mobility law captures the characteristics of dislocation motion in both regimes, with a smooth transition between them. Further, the mobility law captures the dependence of dislocation velocity on stress state, temperature, and local line orientation. In the kink-dominated regime, the free energy of kink-pair nucleation depends on components of stress other than the glide component, hence capturing non-Schmid effects characteristic of bcc plasticity.

The proposed mobility law has a simple analytical form, and its parameters have been obtained from atomistic calculations. A procedure for obtaining the following parameters was described. In particular:

1. Phonon drag coefficients for edge and screw dislocations, and for kinks on screw dislocations have been determined by independent MD calculations. In general, drag coefficients are assumed to posses a temperature-independent contribution (dominant for kinks), and a term proportional to temperature.

2. For screw dislocations, the free energy of kink-pair formation includes both enthalpic and entropic contributions. The enthalpic component is modeled via a Kocks-type function with parameters taken from existing literature, while a constant entropic contribution is determined by matching the flow stress vs temperature profile to experimental data for single-crystal W.

3. A modified Gröger-Vitek model was implemented to take into account non-Schmidt effects. The coefficients of the model were fitted through NEB calculations.

Finally, the mobility law was implemented in DD simulations. First, stress-strain curves for single-crystal W micro-pillars of diameter $\approx 1.1 \mu m$ were obtained for different temperatures and loading directions in tension, in order to verify the effects of the TAT asymmetry of the flow stress. Consistently with experimental evidence, it was found that the flow stress is higher in the anti-twinning orientation. With increasing temperature, it was shown that non-Schmid effects become less pronounced, and that the dislocation microstructure transitions from screw-dominated configuration to a mixed configuration similar to fcc metals and bcc metals at high temperatures, as previously shown by Tang and Marian [22] in DD simulations of Fe crystals. Second, DD simulations were performed to investigate the yield strength of $\langle 001 \rangle$ micro-pillars of different diameters, in both tension and compression. The predicted 0.2% yield strength was found to be consistent with existing experimental values, and the TC asymmetry of the yield stress was captured by the simulation results. Finally, the influence of the loading direction on the flow stress in tension and compression was analyzed. Consistent with experimental measurements conducted on bcc transition metals, the proposed mobility law predicts that the $\langle 011 \rangle$ direction is the hardest in tension, while the $\langle 111 \rangle$ direction is the hardest in compression. Based on these results, it is concluded that the proposed analytical mobility law is well-suited to account for the main features of bcc plasticity in DD simulations.

## Acknowledgements

The authors wish to acknowledge the support of the U.S. Department of Energy, Office of Fusion Energy, through the DOE award number DE-FG02-03ER54708 at UCLA, and the Air Force Office of Scientific Research (AFOSR), through award number FA9550-11-1-0282 with UCLA.